# Simulations on Consumer Tests: Systematic Evaluation of Tolerance Ranges by Model-Based Generation of Simulation Scenarios

**C. Berger**[1], **D. Block**[2], **S. Heeren**[3], **C. Hons**[2], **S. Kühnel**[2], **A. Leschke**[2], **D. Plotnikov**[4] und **B. Rumpe**[4]

[1]Chalmers | University of Gothenburg, Sweden;
[2]Volkswagen AG, Wolfsburg;
[3]Automotive Safety Technologies GmbH, Ingolstadt;
[4]Lehrstuhl für Software Engineering, RWTH Aachen;

**Abstract**

*Context:* Since 2014 several modern cars were rated regarding the performances of their active safety systems at the European New Car Assessment Programme (EuroNCAP). Nowadays, consumer tests play a significant role for the OEM's series development with worldwide perspective, because a top rating is needed to underline the worthiness of active safety features from the customers' point of view. Furthermore, EuroNCAP already published their roadmap 2020 in which they outline further extensions in today's testing and rating procedures that will aggravate the current requirements addressed to those systems. Especially Autonomous Emergency Braking/Forward Collision Warning systems (AEB/FCW) are going to face a broader field of application as pedestrian detection or two-way traffic scenarios.

*Objective:* This work focuses on the systematic generation of test scenarios concentrating on specific parameters that can vary within certain tolerance ranges like the lateral position of the vehicle-under-test (VUT) and its test velocity for example. It is of high interest to examine the effect of the tolerance ranges on the braking points in different test cases representing different trajectories and velocities because they will influence significantly a later scoring during the assessments and thus the safety abilities of the regarding car.

*Method:* We present a formal model using a graph to represent the allowed variances based on the relevant points in time. Now, varying velocities of the VUT will be added to the model while the vehicle is approaching a target vehicle. The derived trajectories were used as test cases for a simulation environment. Selecting interesting test cases and processing them with the simulation environment, the influence on the system's performance of different test parameters will be investigated.

*Results:* The systematic approach reveals several anomalies during the experiments. Due to different test velocities of the VUT within the tolerance ranges the emergency braking guard showed different trigger points which influences an overall rating of a consumer test.

*Conclusion:* The use of the simulation approach allows the systematic evaluation of a black-box algorithm within consumer test scenarios with explicitly defined test parameters by revealing possible anomalies in a system's behavior. Thus, more focused feedback is provided to developers and testers during the development phase. Additionally, real proving ground tests can be planned more efficiently due to reallocating resources on specific aspects.

## 1. Introduction and Motivation

Active safety systems as part of the set of Advanced Driver Assistance Systems (ADAS) are getting more and more in focus of several Consumer-Test-Organizations (CTOs) like IIHS, NTHSA or EuroNCAP. The integration process of such vehicle functions into the compact or small classes as Golf, Polo or up! is ongoing and they are going to be further assessed by those CTOs to establish an indicator for customers regarding the vehicle's safety abilities. It is of special interest for an Original Equipment Manufacturer (OEM) to perform on a high level during these consumer tests to offer the best quality to the customers indicated by a top result. Two development goals are the basis for today's series development: Firstly, to design a system that reacts as quickly as possible on hazardous traffic situations in consumer test scenarios and secondly, to match the customer's expectation regarding their specific driving styles. To achieve both goals a robust parameter setting for the active safety system is required that includes a detailed investigation of possible tolerance ranges of certain test parameters as specified in the EuroNCAP test protocol for example [3].

**Problem Domain and Motivation**

In our first case study [2] we presented a simulation approach to investigate the effects on an emergency braking guard if lateral position and heading of a VUT are varied within certain tolerance ranges. Our experimental case study revealed for a given black box algorithm some anomalies in its behavior for further investigation on real proving grounds for instance. That also helps to reallocate all types of resources more precisely.

**Research Goal and Research Questions**

The research goal for this study is to extend the systematic evaluation method and to provide additional insights regarding the system which would underline the practicability of our approach. For that purpose we investigate the behavior of the emergency braking guard on the example of the "Car-To-Car Rear stationary (CCRs)" by varying the VUT's velocity around the possible trigger point in our simulation environment. The following research questions are of particular interest:

*RQ-1: How can the varying test parameter "velocity" be systematically modeled within certain tolerance ranges while the VUT approaches the target vehicle and to which extent may this parameter influence the trigger points and the VUT's residual velocity?*

*RQ2: How is the residual velocity affected, if the test speed is varied around possible trigger points compared to a constant test speed while the VUT is approaching towards the target vehicle?*

**Contributions of the Article**

This simulation-based approach is extended to model varying velocities of a VUT within the EuroNCAP test scenarios for assessing an emergency braking guard. We could unveil varying trigger points as well as some anomalies due to the nature of changing velocities within the allowed tolerance ranges. Thus, we could show that a simulation environment can significantly support the development and testing process of an active safety system and it enables a more efficient planning and conducting of real world tests.

**Structure of this Article**

In Sec. 2, a selection of related work is briefly illustrated; Sec. 3 presents the main EuroNCAP test scenarios, describes the main boundary conditions of them and gives an outlook of the "roadmap 2020" recently published by EuroNCAP. Afterwards, the simulation environment is shortly presented, the experiment on the example of the CCRs test cases is described and the results are discussed in Sec. 4. The article is summarized and concluded in Sec. 5.

**2. Related work**

The work of Belbachir et al. describes a simulation-based method to evaluate ADAS in different scenarios including the simulation architecture with its environmental and vehicle-based components. They focus on the validation of such systems by considering several self-designed evaluation metrics such as Pedestrian Detections Error for example [4].

During the 2007's DARPA Urban Challenge a simulation environment was part of the software development. The method behind it is presented by Berger and focuses on the automated acceptance testing without the use of real hardware in the first place [5].

Nentwig et al. concentrated explicitly on the use of original hardware delivered by suppliers. They developed a Hardware-in-the-Loop (HiL) testbed that uses synthetically generated data by sensor models to provide corresponding inputs for vision-based ADAS. This approach aims on the support of functional tests during the integration and system testing process ac-cording to the V-model development process using the software tools Virtual Test Drive (VTD) by Vires and the Automotive Data and Time-triggered Framework (ADTF) by Audi Electronic Ventures (AEV) [6][7][8].

Schuldt et al. present the concept of a modularized virtual test tooling kit. Because of the increasing importance of ADAS especially with respect to their later use on public roads intensive testing is mandatory to establish confidence in those vehicle functions to provide the best quality for customers as well as to fulfill legal requirements. Besides the appropriate modeling of vehicle functions they concentrate on the design of assessment criteria in particular to enable a systematic test evaluation approach [9].

Schick et al. also designed a simulation framework for camera- and radar-based ADAS by using a different toolset which is provided by IPG Automotive GmbH. They mainly concentrate on the validation of sensor data fusion algorithms [10]. Moreover, in [11] the use of the vehicle dynamics simulation tool by IPG assessing a chassis control system is presented.

Chucholowski et al. developed a real-time numerical simulation environment modeling the vehicle dynamics for testing virtually passenger cars in the ISO slalom test scenario [12].

Tideman et al. illustrate the toolset "PreScan" by TNO within the evaluation process of a Lane-Keeping Assist (LKA) from a functional point of view [13][14].

## 3. Simulating Consumer Tests

At first, a short summary of EuroNCAP and its AEB test protocol is given. We also briefly outline EuroNCAP's "roadmap 2020" to underline the importance of further investigation of active safety systems as well as further providing simulation methods for analyzing their behavior within these consumer test scenarios. Afterwards, we describe how specific tolerance ranges of the EuroNCAP's AEB test scenarios can be modeled. Furthermore, the used simulation environment is briefly summarized followed by a description of the developed ScenarioDSL, its generator infrastructure and the tools to automate the simulation runs.

**EuroNCAP AEB Test Protocol**

EuroNCAP is a non-profit organization founded and funded by different stakeholders with the purpose to provide detailed information regarding the safety abilities of modern cars for European customers [15].

Since 2014, EuroNCAP aggravated their assessment procedures and defined additional testing scenarios to examine more carefully active safety systems as a subset of ADAS.

Those scenarios are typically derived from databases that recorded crashes occurring in city and inter-urban areas, and which are addressed by AEB/FCW systems as outlined by Fig. 1:

- Car-To-Car-Rear: stationary (CCRs)

- Car-To-Car-Rear: moving (CCRm)
- Car-To-Car-Rear: braking (CCRb)

To evaluate the AEB function within the CCRs scenarios, the vehicle-under-test (VUT) approaches the stationary target vehicle with the desired test velocity that will be increased by 10 km/h in every subsequent test case in a range from 10 km/h to 50 km/h. If a collision between both vehicles occurs, the test speed will be reduced by 5 km/h again and the test run will be continued in 5 km/h steps until the test velocity reaches 50 km/h. If the FCW function is assessed, the aforementioned testing procedure will start with 30 km/h set as test velocity and ends with 80 km/h. During all test scenarios the VUT is driven by a braking and steering robot that reacts on a possible warning of the FCW system by pressing the braking pedal after a certain period of time, defined as "reaction time". The assessment of the AEB function is characterized by a driving robot that gets into a passive mode holding the speed after the test velocity is reached.

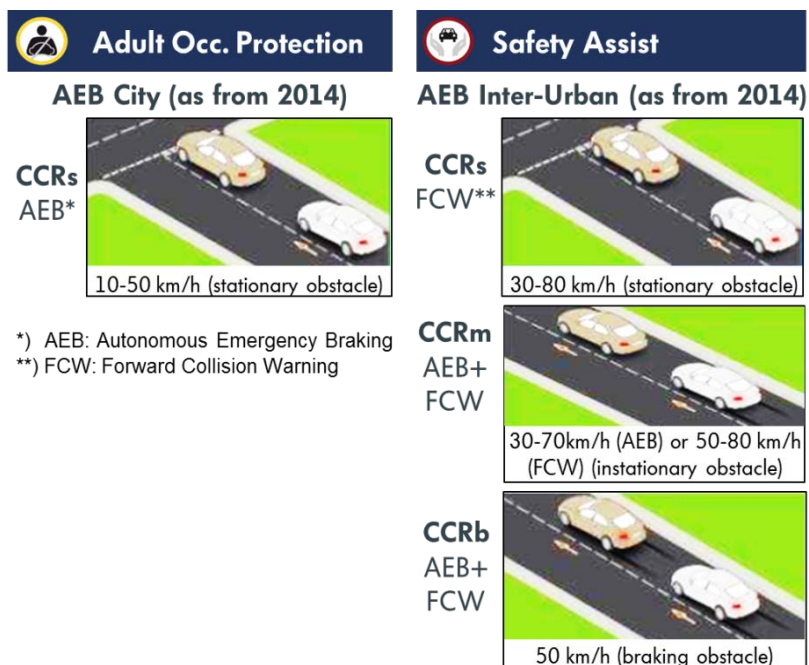

Figure 1: Main scenarios of EuroNCAP's AEB test protocol and the evaluated functions (based on Hulshof et al., 2014 [16]).

Within the CCRm scenarios, the target vehicle is driven with a constant speed of 20 km/h and the VUT's velocity ranges between 30 km/h to 70 km/h for the AEB type and between 50 km/h to 80 km/h for FCW function. The test velocity is increased in the same manner as described before. In the CCRb scenarios, the VUT and the target vehicle are driven with a constant speed of 50 km/h heading in the same direction with a distance of 40 m or 12 m respectively; additionally, the target vehicle decelerates with either 2 m/s² or 6 m/s² [3].

To successfully conduct a test run on a real proving ground, several test parameters must be within particular tolerance ranges as specified in the EuroNCAP AEB test protocol:

- Speed of VUT (test speed + 1.0 km/h)
- Lateral deviation from ideal test trajectory (0 ± 0.1 m)
- Speed of target vehicle (test speed ± 1.0 km/h; only in CCRm and CCRb scenarios)
- Yaw velocity (0 ± 1.0 °/s)
- Steering wheel velocity (0 ± 15.0 °/s)

These parameters must hold between 4s time-to-collision (TTC) and the activation of the active safety system. Otherwise, the test is invalid according to the this test protocol [3].

**EuroNCAP's "2020 Roadmap"**

The new roadmap published by EuroNCAP in June 2014 [1] reveals that current consumer tests are going to be further aggravated due to improvements in the used technology. Thus, additional scenarios will complement the test catalogue by explicitly including vulnerable road users (VRU) like pedestrians, cyclists and powered two-wheelers. Furthermore, scenarios could be possibly added to the current set of scenarios that address different object constellations of oncoming traffic participants turning into or crossing the vehicle's trajectory. Most of these work packages described in [1] might be integrated into the assessments until the year 2020. That fact increases the demand of additional insights derived by simulation approaches, because highly complex requirements have to be met by active safety systems for completing such scenarios successfully. Without those virtual methods, an appropriate evaluation including the different tolerance ranges will hardly be feasible only based on real tests runs.

**Modeling Tolerance Ranges of EuroNCAP Scenarios**

The EuroNCAP AEB test protocol specifies properties of the VUT and means how to properly set up the car. Furthermore, the test protocol defines the points in time where the actual test will start ($T_0$ = TTC = 4s) and under which conditions a test is considered to be invalid: a) When the VUT has left the ideal driving trajectory towards the target vehicle laterally for more than ± 0.1 m including boundaries of yaw and steering wheel velocity, or b) when the VUT has a higher velocity than allowed for the test. The test officially ends, a) when the VUT collides with the target vehicle or b) when the VUT prevent a collision.

To systematically investigate a vehicle's behavior within the tolerance ranges that are allowed for EuroNCAP tests, a model-based representation is required that formally and consistently describes all possible test scenarios. In this regard, our model consists of a graph-based representation where the root of graph $\zeta$ is defined as $T_{End}$ ( = TTC = 0), i.e. that point in time where the VUT collides with the target vehicle. $T_{AEB}$ is the point in time which marks the actual trigger point of the emergency braking guard and is located between $T_0$ and $T_{End}$. The time between $T_0$ and $T_{End}$ is defined as variation period where the VUT is allowed to deviate from the ideal driving trajectory within the given tolerance ranges. This period is shortened to the time interval $[T_0; T_{AEB}]$ if the emergency braking guard is triggered.

The resolution for the variation period describes the number of possible interactions with the VUT to modify its driving behavior; the entire sequence of possible interactions describes a possible path in the graph ζ that also defines the height of ζ. All possible paths p of ζ describe all potential interactions that can be applied to the VUT; the representation of all paths p originating from $T_{End}$ towards $T_0$ describes an exponentially growing tree. Since it is practically not possible to generate and simulate all paths p, appropriate algorithms are required to select an important and representative subset thereof.

**The Simulation Environment**

The simulation environment is set up with its main tools VTD by Vires and ADTF by Audi Electronic Ventures. VTD provides a broad mass of static and dynamic context data unified in the "Runtime Data Bus" (RDB); both the static context like roads, houses, traffic signs for instance, as well as the dynamic one like vehicles and their actions, pedestrians and other types of moving objects for example can be placed within VTD by editors and saved as scenarios in an XML-based notation [17][18].

An RDB message is sent via a TCP/IP interface to ADTF, in which the actual system is modelled with its different software components, i.e. sensors, algorithms and actuators. Additional components ensure necessary data transformations to assure valid input formats for black box components. Selected ADTF configuration data streams are plotted over time in a comma-separated-value (CSV) file similar to the format as it is recorded during real vehicle tests; afterwards, these CSV files are used for further evaluation purposes. For a more detailed description of the general architecture of the environment, it is referred to [19].

**A ScenarioDSL and the ScenarioGenerator**

In order to analytically map the derived scenario definitions, which describe EuroNCAP test scenarios and concrete paths that model allowed variations as described in section B into our simulation environment, we developed a Domain Specific Language (DSL) termed as ScenarioDSL in the remainder.

The purpose of the ScenarioDSL generator is to generate an XML-based representation for the scene in VTD from an abstract scenario description in ScenarioDSL and to initialize the virtual driver and the vehicle dynamics with concrete configuration and path data. The ScenarioDSL was developed with the domain specific language workbench MontiCore [20], [21],

[22]. MontiCore enables a definition of textual DSLs based on an extended context free grammar format and supports mechanisms for language composition like language inheritance, embedding and aggregation as presented in [23]. The extended grammar format is used to derive an Abstract Syntax Tree (AST) and parsers that process textual DSL instances and store their representation as an AST instance [24]. Finally, MontiCore facilitates the Freemarker Template engine [25] to generate code or textual models for the target system [26], e.g. XML- based scenarios for the VTD. Furthermore, MontiCore is able to generate Eclipse-based editors with support of the syntax highlighting, folding, outline, and code completion [27].

```
scenario CCRs_Base {
  Layout {Database = "test.xodr"}
  VehicleList {
     ConfigFile = "cfg.xml" }
  ...
  TrafficElements {
      Player VehicleUnderTest{
      Description {
        Driver=DefaultDriver
        Control=external
        Type=Brand_VehicleProject
      }

  }
  PulkTraffic { ... }
  ...
}
```

*Figure 2: Extract of the base scenario for CCRs case study.*

```
scenario CCRs_25kmh extends
CCRs_Base {

  TrafficElements {
      Player VehicleUnderTest{
      Init {
        PosAbsolute = (0,0,0,true)
      }

  }
}
```

*Figure 3: Compact concrete scenario description for the 25 km/h test case.*

The ScenarioDSL is conceptually derived on the XSD-definition of scenarios in VTD and, thus, possesses the same expressiveness. Furthermore, it offers abstraction mechanisms like inheritance and overriding of certain elements known from the Object Oriented programming languages [28]. This enables concise definition of scenarios in EuroNCAP context, where different scenarios differ only in dedicated places, e.g. different geometric positions of the VUT.

Fig. 2 represents an extract of the scene definition; and Fig. 3 shows the entire scene definition. The generator for the ScenarioDSL merges these DSL instances into a complete scene definition for the VTD.

**Automating Simulation Runs**

In order to automate the simulation workflow, we extended our approach presented in [2] with a controlling component based on a versioning server and VTD simulation controller.

Since it is important to track how particular algorithm's parameter values influence the EuroNCAP performance, we store the vehicle's traces for each test case together with the test case's configuration in the versioning server Subversion (SVN).

As shown in Fig. 4, the modules described in [2] are logically embedded into a controlling component termed as a SVN Automation Module. This component is composed of two major parts: an SVN server and observer on the one side, and the SCP based controller on the other side. In order to control the simulation execution, a further ADTF filter was developed. This filter observes incoming additions on the server, which are considered as simulation jobs, and starts and stops simulation in the VTD environment after the evaluation of a particular test scenario is expected to be accomplished.

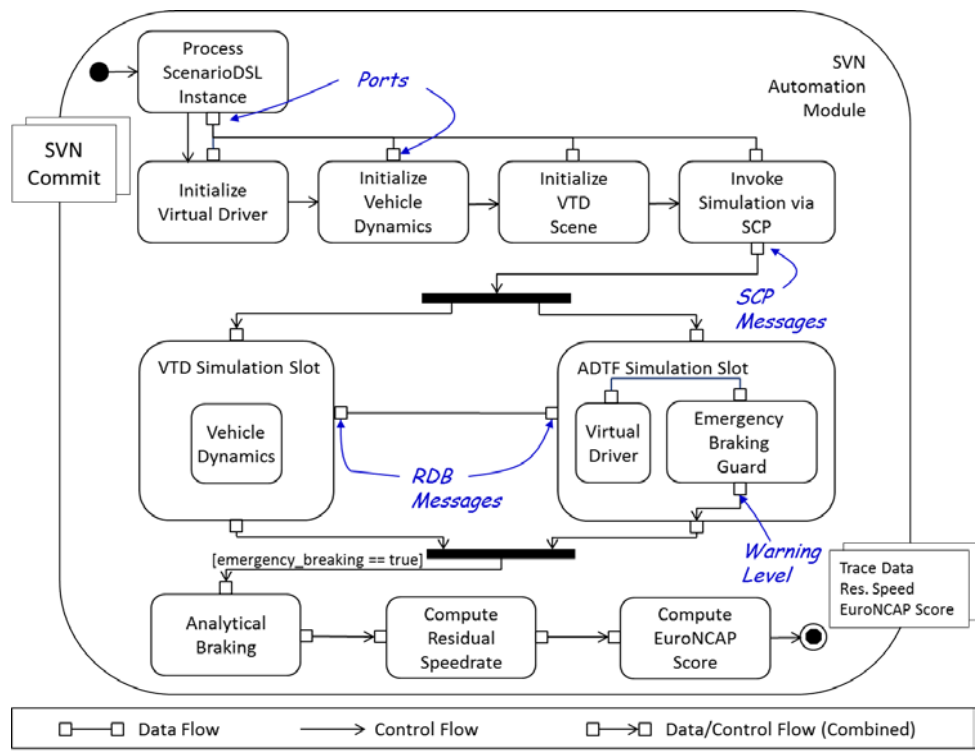

*Figure 4: Simulation workflow for the presented method.*

The simulation of a test case is started as soon as a new ScenarioDSL instance is uploaded to the SVN server. After that, the ScenarioDSL generator processes this instance to produces an XML-based scene description and to initialize the concrete paths for vehicle dynamics and virtual driver. The virtual driver module implemented as an ADTF filter simulates an EuroNCAP testing robot. The vehicle module serves for precise positioning of the VUT in the test scene. Afterwards, the simulation is executed and the produced VUT traces are stored on the SVN server for the subsequent evaluation and scoring calculation. Finally, if there are further unprocessed instances on the SVN server, the process will be repeated for them; otherwise, the simulation environment idles in its current state.

## 4. Empirical Case Study

In the following, we describe our experimental case study on considering velocity variations during the systematic evaluation of tolerance ranges for the EuroNCAP CCRs AEB 25 km/h test case. We report according to guidelines from Ledlitschka et al. [25] and Shull et al. [26].

**Experimental Setup**

The goal of the experiment is to systematically analyze how a black-box active safety function handles a traffic situation that becomes more critical just before the action point in time. Here, it is of interest for an OEM to validate whether requirements for an algorithm to adequately react on increasing criticalities are properly realized by a supplier.

The experiment is defined as following: The VUT and target vehicle are located 67.5 m apart from each other. The target vehicle has a velocity of $v_{target}$ = 0 m/s throughout the entire experiment.

[Exp-1]: "Constant speed" as reference runs:

*To determine the behavior of the VUT, a reference run is conducted where the velocity of VUT is not modified.*

The VUT will accelerate with $a_{VUT}$ = 2 m/s² until its target velocity of 25.8 km/h is reached. The following lateral variations are applied to the VUT: a) No lateral deviation; b) oscillating lateral deviation up to 0.1 m starting at the left hand side of the ideal trajectory; and c) oscillating lateral deviation up to 0.1 m starting at the right hand side of the ideal trajectory.

[Exp-2]: "Speed variation" as analysis runs:

*Subsequently, the experiment is repeated by using the action point in time $T_{AEB}$ that was determined in the previous runs. Additionally, the behavior of the VUT is adapted so that at the action point in time, a higher velocity is achieved.*

The VUT will accelerate with $a_{VUT}$ = 2 m/s² until its test velocity of 25 km/h is reached. It will accelerate with $a_{VUT,modify}$ = 0.1 m/s² at ($T_{AEB} - T_{initiate}$) until its test velocity of 25.4 km/h and 25.8 km/h is reached, respectively. The lateral variations are applied to the VUT in the same manner as described in [Exp-1]: "Constant speed". The point in time ($T_{AEB} - T_{initiate}$) for experiment [Exp-2] is determined empirically.

**Experimental Procedure**

The experimental procedure is described briefly in the following. After initializing the scene in VTD, the configuration of the virtual driver and the vehicle dynamics according to the path p, the simulation is started by an SCP command. During the simulation, the VUT's trace data is collected and after the SCP controller stopped the simulation run, the data is stored on the SVN server. Afterwards, the expected $v_{res}$ is analytically determined and stored alongside with the VUT's trace on the versioning server SVN. The data is sufficient to compute the EuroNCAP score.

**Experimental Results**

In this subsection the results from [Exp-1]: "Constant speed" and [Exp-2]: "speed variation" are presented.

| Test case | "25.0 km/h" | | | "25.4 km/h" | | | "25.8 km/h" | | |
|---|---|---|---|---|---|---|---|---|---|
| Oscillation and experiment | $v_{AEB}$ [km/h] | $D_x$ [m] | $v_{res}$ [km/h] | $v_{AEB}$ [km/h] | $D_x$ [m] | $v_{res}$ [km/h] | $v_{AEB}$ [km/h] | $D_x$ [m] | $v_{res}$ [km/h] |
| Exp-1 Left | 25.0 | 8.72 | 4.8 | 25.4 | 9.36 | 1.81 | 25.8 | 9.43 | 3.76 |
| Exp-1 Ideal | 25.0 | 9.28 | 0.0 | 25.4 | 9.08 | 5,40 | 25.8 | 9.14 | 6.35 |
| Exp-1 Right | 25.0 | 8.72 | 4.8 | 25.4 | 9.36 | 1.81 | 25.8 | 9.43 | 3.76 |
| Exp-2 Left | 25.0 | 8.72 | 4.8 | 25.35 | 9.53 | 0.0 | 25.77 | 9.38 | 2.88 |
| Exp-2 Ideal | 25.0 | 9.28 | 0.0 | 25,4 | 9.23 | 0.0 | 25,77 | 9.93 | 0.0 |
| Exp-2 Right | 25.0 | 8.72 | 4.8 | 25.35 | 9.53 | 0.0 | 25,77 | 9.38 | 2.88 |

*Figure 5: Results of [Exp-1] and [Exp-2] with trigger point $T_{AEB}$, distance $D_x$ to the target vehicle and the residual speed $v_{res}$.*

[Exp-1]: Fig. 5 illustrates the varying distances $D_x$ of the VUT to the target vehicle for each trajectory with a constant velocity, after the emergency braking guard sent a signal to brake. The residual velocity $v_{res}$ is calculated by defining an analytical braking function over time with a deceleration rate of a = 3.5 m/s² and a delay of 0.3 s, estimating the according TTC at $T_{AEB}$. Moreover, $v_{AEB}$ shows the actual velocity of the VUT at the same point in time.

[Exp-2]: In Fig. 5, the distances $D_x$ between the VUT and the target vehicle are also shown for speed variation. In this case, the velocity is increased by 0.1 m/s² until the desired velocity of 25.4 km/h and 25.8 km/h is reached, respectively. The point in time when to increase the velocity bases on the trigger point of the emergency braking guard from the 25.0 km/h test case.

**Discussion and Analysis**

In the following the experimental results from Exp-1 and Exp-2 are analyzed and discussed. Fig. 6 illustrates the general test scenario with a test speed of 25 km/h and an additional variation from the tolerance range.

[Exp-1]: Fig. 5 reveals that in case of the ideal trajectory of the "25.0 km/h" and the "25.4 km/h" test case a first anomaly in the behavior of the active safety system could be recog-

nized because $D_x$ between the two vehicles decreases first, but then starts to increase again in the case of the "25.8 km/h" test case. Furthermore, in case of "25.4 km/h" and "25.8 km/h" tests, a lateral deviation by an oscillating trajectory results in an earlier triggering of the system compared to the ideal trajectory; in case of the "25.0 km/h", test this trend is turned around, which is the actual result that we expected. These anomalies in the behavior of the black-box algorithm of the emergency braking guard could be caused by several influences and need to be discussed between the test engineers at the OEM and the supplier of the algorithm.

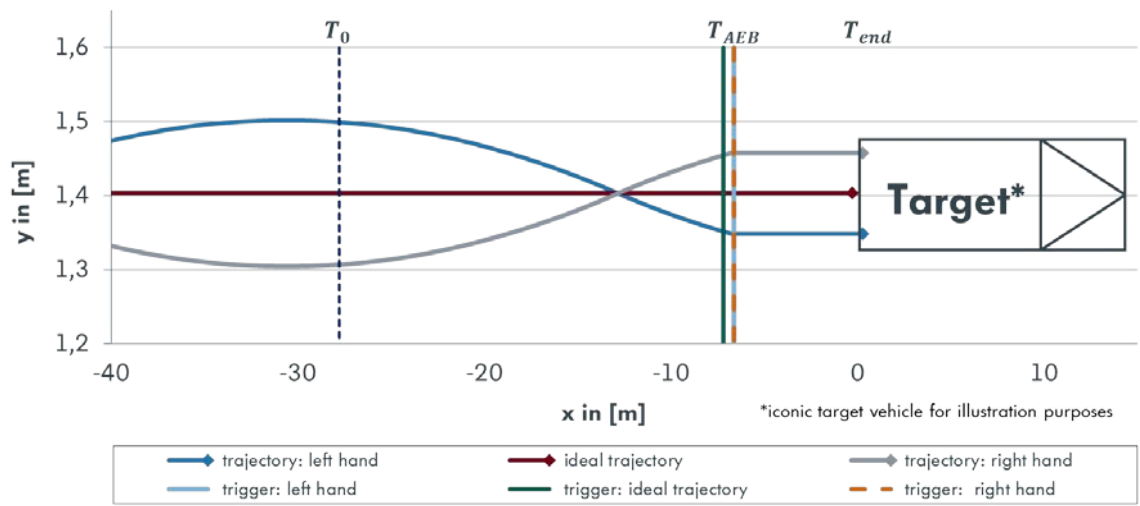

*Figure 6: General scheme of the 25 km/h test scenario as a basis for the experiments.*

[Exp-2]: The anomaly of inverted trigger points continues between the different test cases with a constant test velocity and the ones with varying speeds. Because of the acceleration of the VUT, the emergency braking guard starts acting earlier compared to the test cases with constant speed. Moreover, during the VUT's approach the desired test velocity is not fully reached which fact results in a lower residual speed of the VUT later on. Furthermore, in case of a varied "25.4 km/h" test case a collision could be avoided based on our assumptions. In addition, every test case that is performed with varying velocities and the VUT's trajectory is ideal; a collision could be avoided as well.

**Threats to Validity**

Hereafter, we are reporting about the threats to validity to our study according to the guidelines provided by Runeson and Höst [27].

Regarding construction validity, the goal of the experiment was to systematically analyze the behavior of the VUT in the case that a given traffic situation becomes more critical right before the expected action point in time. Therefore, we conducted two experiments: The first one is the control experiment as the velocity of the VUT is not modified throughout the EuroNCAP experiment. In the second experiment, we modify the VUT's velocity right before the expected time point of action as the only parameter according to the allowed tolerance range. Thus, we can compare the behavior of the VUT and hence, the experiment setting allows for this comparison of interest.

Regarding internal validity, the experiments are planned and conducted by using officially available EuroNCAP test protocols. Therefore, specific procedures that might favor the considered VUT can be ruled out.

Regarding external validity, our study was conducted in a virtual environment. While the components comprising the simulation environment are used during the development of components for the VUT and thus, the suitability of the environment for such studies can be confirmed, the results need to be confirmed with selected experiments on proving ground to validate the observation. Moreover, further studies are necessary to identify a trend in the observed behavior in the simulation for increasing velocities for example, and to transfer the results in general.

## 5. Conclusion

In our work we extended our simulation-based approach to systematically evaluate the influence of tolerance ranges for the EuroNCAP tests based on an emergency braking guard system. We conducted two experiments to examine the influence of the test velocity within one particular test scenario to investigate to which extent the trigger points and the residual speed vary.

Addressing our first research question, [Exp-1] showed that the trigger points changed within the officially allowed tolerance ranges of the EuroNCAP AEB test protocol. Some anomalies were also unveiled due to inverted trigger points of the underlying system; further analysis with experts from the supplier is needed as well as performing additional real test runs on a proving ground. [Exp-2], which addresses our second research question, revealed lower residual velocities for the VUT with increased test speeds close to a possible trigger point of the system compared to those tests with constant test speed.

This simulation-based approach allows a systematic analysis of an active safety system under development during the testing phase. By identifying possible anomalies in critical test cases, real world test runs can be planned, conducted and assessed more efficiently due to additional insights by the simulation. Future work will concentrate on improvements of the underlying models for the different system components as well as considering the upcoming challenges derived from EuroNCAP's roadmap 2020.